\documentclass[preprint,preprintnumbers,amsmath,amssymb]{revtex4}

\usepackage{graphicx}% Include figure files
\usepackage{dcolumn}% Align table columns on decimal point
\usepackage{bm}% bold math

\begin{document}

%\preprint{APS/123-QED}

\title{Equivalence Principle and the Gauge Hierarchy Problem}% Force line breaks with \\

\author{Xavier Calmet}
 \email{xavier.calmet@ulb.ac.be}
 \affiliation{Universit\'e Libre de Bruxelles, 
Service de Physique Th\'eorique, CP225,
Boulevard du Triomphe  (Campus plaine),
B-1050 Brussels, Belgium. 
}%

\date{\today}% It is always \today, today,

\begin{abstract}
 We show that the gauge hierarchy problem can be solved in the framework of scalar-tensor theories of gravity very much in the same way as it is solved in the Randall-Sundrum scenario. Our solution involves a fine-tuning of the gravitational sector which can however be avoided if a supergravity extension of the dilaton sector is considered. However our mechanism does not require the introduction of extra-dimensions or new physics strongly coupled to the standard model in the low energy regime. We do introduce a new scalar field which is however coupled only gravitationally to regular matter. The physical reason for the splitting between the weak scale and the Planck scale is a violation of the Einstein's equivalence principle. 
\end{abstract}

%\pacs{ blabla}% PACS, the Physics and Astronomy
\maketitle

It is widely believed that a  stabilization of  the scale of the  electroweak symmetry spontaneous breaking requires the introduction of some new physics around the weak scale which couples strongly to the standard model. A textbook example are supersymmetric extensions of the standard model (see e.g. \cite{Nilles:1983ge} for a review), where in order to solve the hierarchy problem, the scale for supersymmetry breaking cannot decouple from the weak scale and hence the couplings between the standard model particles and their superpartners have to be sizable. Another example are technicolor models (see e.g. \cite{Farhi:1980xs} for a review) which predict a plethora of new particles.  More recently the idea that extra-dimensions  \cite{ArkaniHamed:1998rs,Randall:1999ee} could address the hierarchy problem has attracted a lot of attention. The LHC is hence expected to discover either new particles or effects of extra-dimensions. Here we give a counter example and show that the scale of electroweak symmetry breaking can be stabilized by Planck scale physics effects and that no new physics interacting strongly with the standard model at the weak scale is necessary to stabilize the weak scale. The physical reason for the splitting between the weak scale and the Planck scale is a violation of the equivalence principle.

In order to establish our notations, we shall first review briefly how to transform the Jordan-Brans-Dicke action in the Jordan frame to the corresponding action in the Einstein frame. 
The  Jordan-Brans-Dicke \cite{Jordan,Brans:1961sx}  action 
  \begin{eqnarray}
  S_{JBD}&=&  \int d^4x \sqrt{-\hat g} \left ( \frac{1}{8} \hat \phi^2 \hat R - \omega \frac{1}{2} \hat g^{\mu\nu} \partial_\mu \hat \phi \partial_\nu \hat \phi \right )
\end{eqnarray}
can be mapped to the Einstein frame \cite{Dicke:1961gz} (see also \cite{Schmidt:1988xi,Dick:1998ke,Faraoni:1998qx,Faraoni:1999hp} for more recent papers on the topic)
 \begin{eqnarray}
   S_{E}&=&  \int d^4x \sqrt{-g} \left ( \frac{1}{2 \kappa}   R - \frac{1}{2}  g^{\mu\nu} \partial_\mu  \phi \partial_\nu  \phi \right )
\end{eqnarray}
using
 \begin{eqnarray}
   \hat \phi = 2 m_p \exp \left (  \frac{ \sqrt{2} }{2 \sqrt{3+ 2 \omega}} \frac{\phi}{m_p} \right )
\end{eqnarray}
and 
 \begin{eqnarray}
   \hat g_{\mu\nu} =  \exp \left (- \frac{ \sqrt{2} }{\sqrt{3+ 2 \omega}} \frac{\phi}{m_p} \right )  g_{\mu\nu}
\end{eqnarray}
 with $\omega>-3/2$ and $\kappa=8 \pi G$.

Let us introduce the following notations
  \begin{eqnarray}
  \Omega(x)^2= \exp \left (-\frac{ \sqrt{2} }{\sqrt{3+ 2 \omega}}  \frac{\phi}{m_p} \right )
 \end{eqnarray}
 with $m^2_p=1/(8 \pi G)$ i.e. the reduced Planck mass. Note that the Planck scale is $\Lambda_p^2=1/G$. Under a Weyl transformation one has (see e.g. \cite{Birrell:1982ix}):
  \begin{eqnarray}
 \hat R= \Omega^{-2}(x) R - 6 \Omega^{-3}(x) \Omega_{;\mu\nu}(x) g^{\mu\nu}
  \end{eqnarray}
 \begin{eqnarray}
\sqrt{- \hat g} = \Omega^4(x) \sqrt{- g}
  \end{eqnarray}
  and the scalar field transforms according to (note that this is not a Weyl transformation)
   \begin{eqnarray}
\hat \phi = \Omega^{-1}(x) 2 m_p.
  \end{eqnarray}
 Let us now consider the Higgs sector of the standard model of particle coupled to the Jordan-Brans-Dicke action and we add a potential for the Jordan-Brans-Dicke field (JBD-field) as well as a cosmological constant. We will denote the Higgs doublet by H:
  \begin{eqnarray}
  S_{JBD}&=&  \int d^4x \sqrt{-\hat g} \left ( \frac{1}{8} \hat \phi^2 \hat R - \frac{\omega}{2} \hat g^{\mu\nu} \partial_\mu \hat \phi \partial_\nu \hat \phi - \frac{1}{2} m_{\hat \phi}^2 \hat \phi^2 - \frac{\lambda_{\hat \phi}}{4} \hat \phi^4 -2 \Lambda 
   \right .
    \\  &&  \nonumber  \left .
 - \frac{1}{2}  \hat g^{\mu\nu} (D_\mu  H)^\dagger D_\nu  H -\mu^2 H^\dagger H - \lambda (H^\dagger H)^2 + \lambda_2 H^\dagger H \hat \phi \hat \phi +\alpha H^\dagger H \hat \phi \right ).
\end{eqnarray}
The potential for the scalar field which couples to the Ricci scalar, which we call the JBD-scalar, and in particular the mass term prevents  conflicts with experiment since our scalar field will not lead to a fifth force type interaction as it is the case in the  Jordan-Brans-Dicke gravity. In particular if  the mass of the JBD-scalar field  is  large,  it does not propagate much.   Let us now look at this theory in the Einstein frame, one gets
 \begin{eqnarray}
   S_{E}&=& \int d^4x \sqrt{-g} \Big ( \frac{1}{2 \kappa}   R
     - \frac{1}{2}  g^{\mu\nu} \partial_\mu  \phi \partial_\nu  \phi
      - 2  m_{\hat \phi}^2   m_p^2   \Omega^{2}(x) 
      - 4 \lambda_{\hat \phi} m_p^4
      \\  &&  \nonumber
      -2 \Lambda  \Omega^4(x)
   - \frac{1}{2}  \Omega^2(x) g^{\mu\nu} (D_\mu  H)^\dagger D_\nu  H -  \Omega^4(x) \mu^2  H^\dagger H   - \lambda  \Omega^4(x) (H^\dagger H)^2 
      \\  &&    \nonumber
       + 
         \lambda_2 4 m^2_p H^\dagger H  \Omega^{2}(x) 
        +  \alpha 2 m_p H^\dagger H \Omega^3(x) \Big ).
    \end{eqnarray}
    Let us now assume that the Higgs boson acquires a vacuum expectation value $v=\sqrt{-\mu^2/(2\lambda)}$. Using the unitary gauge, the action becomes
    \begin{eqnarray}
   S_{E}&=& \int d^4x \sqrt{-g} \Big ( \frac{1}{2 \kappa}   R
     - \frac{1}{2}  g^{\mu\nu} \partial_\mu  \phi \partial_\nu  \phi
      - 2  m_{\hat \phi}^2   m_p^2   \Omega^{2}(x) 
      - 4 \lambda_{\hat \phi} m_p^4
      \\  &&  \nonumber
      -(2 \Lambda - \lambda v^4) \Omega^4(x)
   - \frac{1}{2}  \Omega^2(x) g^{\mu\nu} \partial_\mu  h \partial_\nu  h - 4  \Omega^4(x) \lambda v^2  h^2  - 4  \Omega^4(x) \lambda v h^3
      \\  &&    \nonumber
        -  \lambda    \Omega^4(x) h^4  + 
         \lambda_2 4 m^2_p (h+v)^2  \Omega^{2}(x) 
        +  \alpha 2 m_p (h+v)^2 \Omega^3(x) \Big ).
    \end{eqnarray}
where we have discarded the terms involving the gauge bosons of the standard model which however do not introduce any further complication.
     The vacuum expectation value $v$ of the Higgs field is  expected to be large because of radiative corrections and typically of the order of the Planck scale if this scale is used to regularize the quadratically divergent contributions to the Higgs squared mass. This is the gauge hierarchy problem. 
 Let us rescale the Higgs field using $h \to   \Omega^{-1}(x) h$, we get
 \begin{eqnarray} \label{fundaction}
   S_{E}&=& \int d^4x \sqrt{-g} \Big ( \frac{1}{2 \kappa}   R
     - \frac{1}{2}  g^{\mu\nu} \partial_\mu  \phi \partial_\nu  \phi 
     \left ( 1+ \frac{h^2}{2 (3+2 \omega) m_p^2} \right ) 
      \\  &&  \nonumber
      - 2  m_{\hat \phi}^2   m_p^2    \Omega^{2}(x) 
      - 4 \lambda_{\hat \phi} m_p^4
      -(2 \Lambda -\lambda v^4)  \Omega^4(x)
      \\  &&  \nonumber
   - \frac{1}{2}  g^{\mu\nu} \partial_\mu  h \partial_\nu h - \frac{\Omega(x) \sqrt{2}}{2\sqrt{3+2\omega}}  \frac{h}{m_p}  g^{\mu\nu} \partial_\mu  \phi \partial_\nu h 
     - 4  \Omega^2(x) \lambda v^2  h^2  - 4 \Omega(x) \lambda v h^3
      \\  &&    \nonumber
        -  \lambda    h^4 
         +  \lambda_2 4 m^2_p (h+v \Omega(x))^2  
        +  \alpha 2 m_p (h +v \Omega(x))^2 \Omega(x)
         \Big ).
    \end{eqnarray}
    We see that the vacuum expectation of the Higgs field, which determines the electroweak symmetry breaking scale, is corrected by an exponential function. We will argue that we have to set $\alpha$ and $\lambda_2$ to zero, since $\hat \phi$ is not gauged, this can be done consistently at all renormalization scales. In other words the operators $ \lambda_2 H^\dagger H \hat \phi \hat \phi$ and $\alpha H^\dagger H \hat \phi$ will not be generated by radiative corrections.
    
Note that the terms  $2m^2_{\hat \phi} m_p^2 \Omega^2(x)$ and $(2\Lambda-\lambda v^4) \Omega^4(x)$  contain mass terms for the scalar field $\phi$ as well as self-interacting terms which are highly non-linear.  We are assuming that parameters of the action are such that the field $\phi$ develops a vacuum expectation value:
\begin{eqnarray}
  \Omega(x)= e^{- \frac{\sqrt{2}}{2 \sqrt{3+2 \omega}} \frac{\xi(x)+v_\xi}{m_p}}=
  e^{- \frac{\sqrt{2}}{2 \sqrt{3+2 \omega}} \frac{\xi(x)}{m_p}} 
  e^{- \frac{\sqrt{2}}{2 \sqrt{3+2 \omega}} \frac{v_\xi}{m_p}}
  = \bar \Omega(x) \Omega.
  \end{eqnarray}
The scalar potential for the JBD-field is given by
\begin{eqnarray}
 V=2 m_{\hat \phi}^2 m_p^2 \Omega^2(x) + (2 \Lambda-\lambda v^4) \Omega^4(x).
   \end{eqnarray}
The first derivative of this potential with respect to the JBD-field  is given by
\begin{eqnarray}
\frac{\partial V}{\partial \phi}=4 \left ( m_{\hat \phi}^2 m_p^2  + (2 \Lambda-\lambda v^4) \Omega^2(x) \right) \Omega(x) \frac{\partial  \Omega}{\partial \phi}(x).
   \end{eqnarray}
   The non trivial minimum of this potential is at
   \begin{eqnarray}
\phi_{min}= - m_p \sqrt{\frac{3+2\omega}{2}} \ln \left ( -\frac{m_{\hat \phi}^2 m_p^2}{2 \Lambda-\lambda v^4} \right).
   \end{eqnarray}
Naturalness forces us to assume that $v\sim m_p$, $|\Lambda| \sim m_p^4$ and $\lambda \sim {\cal O}(1)$. In orther words, $ |2 \Lambda-\lambda v^4| \sim |\Lambda| \sim m_p^4$, we thus see that either $m_{\hat \phi}^2$ or $\Lambda$ has to be negative. Furthermore in order to explain the hierarchy of the scales,  we need to have a vacuum expectation value which is of the order of the Planck scale i.e.
\begin{eqnarray}
  \Omega=\left ( -\frac{m_{\hat \phi}^2 m_p^2}{2 \Lambda-\lambda v^4} \right)    \sim 10^{-17},
 \end{eqnarray}
 which implies that $m_{\hat \phi}\sim 1 \times 10^{10.5}$ GeV assuming that the Planck scale if of the order $10^{18}$ GeV. Note that this fine tuning is in the gravitational sector since it is a fine tuning of the mass of the of the JBD-scalar which fixes the Planck scale. This parameter might be fixed to the desired value by quantum gravitational effects. It would be important to have a stabilization mechanism similar to the Goldberger-Wise mechanism \cite{Goldberger:1999uk}. There are two contributions to the mass of the JBD-fields which destabilize the scale $m_{\hat \phi}$. The first class of corrections come from the self-interactions of the scalar field. It is easy to see that a supersymmetric extension of this sector can stabilize this mass scale. We introduce a new singlet fermion which is assumed to be the superpartner of the dilaton. It's mass is assumed to be of the order of $m_{\hat \phi}$ and  supersymmetry is softly broken to avoid spoiling the cancellation  of quadratic divergences to the JBD-scalar quadratic mass.
 
 Another class of contribution to the JBD-scalar quadratic mass are from quantum gravitons. The coupling $\sqrt{-g} \phi^2 R$ naively can lead to large contributions to the dilaton like field of the order of $M_p^2$ if the cutoff is assumed to be the Planck mass. Here again, it is easy to stabilize this scale using supergravity. The gravitino will cancel the contributions of the graviton to the JBD-field. The gravitino is assumed to have a mass of the order of that of the JBD-field.  We assume that supersymmetry is completely broken in the standard model sector.

 It is important to notice that the value of the Jordan-Brans-Dicke parameter $\omega$ does not impact the solution to the gauge hierarchy problem. Let us now study the second derivative of the potential evaluated at $\phi_{\min}$, we get:
 \begin{eqnarray}
m_\phi^2= \frac{\partial^2 V}{\partial \phi \partial \phi} \vline_{_{_{_{_{_{_{_{_{_{\phi=\phi_{\min}}}}}}}}}}}=\frac{4 m_{\hat \phi}^2}{3+2 \omega} \Omega^2(\phi_{\min})+ \frac{16 \Lambda-8 \lambda v^4}{m_p^2(3+2\omega)} \Omega^4(\phi_{\min})
   \end{eqnarray}
 which has to be positive. We hence pick $\Lambda<0$, $m_{\hat \phi} \sim 10^{10.5}$ GeV  and $\sqrt{|\Lambda|} \sim m_p^2$.
Note that the mass of the JBD-scalar in the Einstein frame is of the order of the $m_{\phi}=10^{-13}$ GeV and is thus extremely light.
 
 For the parameter range we have chosen we can expand the action in $1/m_p$ and obtain:
 \begin{eqnarray}
   S_{E}&=& \int d^4x \sqrt{-g}   \Big (   \frac{1}{2 \kappa}   R
     - \frac{1}{2}  g^{\mu\nu} \partial_\mu  \xi \partial_\nu  \xi 
     \left ( 1+ \frac{h^2}{2 (3+2 \omega) m_p^2} \right )  
      \\  &&  \nonumber 
      - 2  m_{\hat \phi}^2   m_p^2 \Omega^{2} 
       f_2(\xi)
                  - 4 \lambda_{\hat \phi} m_p^4
      -(2 \Lambda -\lambda v^4)   \Omega^4
       f_4(\xi) 
      \\  &&  \nonumber
   - \frac{1}{2}  g^{\mu\nu} \partial_\mu  h \partial_\nu h -
    \frac{ \sqrt{2}\Omega}{2\sqrt{3+2\omega}}  \frac{h}{m_p} 
    g^{\mu\nu} \partial_\mu  \xi \partial_\nu h   f_1(\xi)    
      - 4   \lambda v^2  h^2  \Omega^{2} 
       f_2(\xi)   
       - 4 \lambda v h^3
      \Omega
      f_1(\xi)  
      \\  &&    \nonumber 
        -  \lambda    h^4  
        +
          \lambda_2 4 m^4_p \left (h+ v    \Omega f_1(\xi)   \right)^2  +  \alpha 2 m_p \left (h  +v  \Omega  f_1(\xi) \right )^2  \Omega f_1(\xi) \Big)
         \end{eqnarray}
    where $f_1(\xi)$,  $f_2(\xi)$ and $f_4(\xi)$ are given respectively by:
\begin{eqnarray}
f_1(\xi)&=&   \left (1- \frac{\sqrt{2}}{ 2 \sqrt{3+2\omega}} \frac{\xi}{m_p}
      +  \frac{1}{4 (3+2\omega)} \frac{\xi^2}{m^2_p} \right ) + {\cal O}\left(\frac{\xi}{m_p}\right)^3, \\
f_2(\xi)&=& \left (1- \frac{\sqrt{2}}{  \sqrt{3+2\omega}} \frac{\xi}{m_p}
      +  \frac{1}{(3+2\omega)} \frac{\xi^2}{m^2_p} \right )+ {\cal O}\left(\frac{\xi}{m_p}\right)^3,
      \\
  f_4(\xi)&=&  \left (1- \frac{2\sqrt{2}}{  \sqrt{3+2\omega}} \frac{\xi}{m_p}
      +  \frac{4}{(3+2\omega)} \frac{\xi^2}{m^2_p} \right )+ {\cal O}\left(\frac{\xi}{m_p}\right)^3.
\end{eqnarray}
    
We see that for our mechanism to work, two conditions need to be fulfilled. The JBD-scalar must couple only gravitationally to the rest of matter and we thus impose $\lambda_2=0$ and $\alpha=0$. This is not really unnatural since the JBD-scalar field is not gauged and hence the constants $\lambda_2$ and $\alpha$ will not get renormalized  by quantum effects of the gauge sector and can thus be set to zero at all scale as long as quantum gravitational interactions are discarded. At first sight one may worry that quantum gravity could reintroduce these terms however the operator $h^\dagger h  \phi$ cannot be generated by quantum gravity loops and the operator $h^\dagger h \phi^2$ is suppressed by the factor $m^2_h m^2_{\phi}/\Lambda_p^4$ and is thus not a problem.  The low energy action is thus:
 \begin{eqnarray}
   S_{E}&=& \int d^4x \sqrt{-g}   \left (   \frac{1}{2 \kappa}   R
     - \frac{1}{2}  g^{\mu\nu} \partial_\mu  \xi \partial_\nu  \xi 
     \left ( 1+ \frac{h^2}{2 (3+2 \omega) m_p^2} \right )  \right . 
      \\  &&  \nonumber \left .
      - 2  m_{\hat \phi}^2   m_p^2 \Omega^{2} 
       \left (1- \frac{\sqrt{2}}{ \sqrt{3+2\omega}} \frac{\xi}{m_p}
      +  \frac{1}{(3+2\omega)} \frac{\xi^2}{m^2_p} \right )  \right .
       \\  &&  \nonumber  \left .
                  - 4 \lambda_{\hat \phi} m_p^4
      -(2 \Lambda -\lambda v^4)   \Omega^4
        \left (1- \frac{2\sqrt{2}}{ \sqrt{3+2\omega}} \frac{\xi}{m_p}
      +  \frac{4}{(3+2\omega)} \frac{\xi^2}{m^2_p} \right )  \right .
      \\  &&  \nonumber \left .
   - \frac{1}{2}  g^{\mu\nu} \partial_\mu  h \partial_\nu h -
    \frac{ \sqrt{2}\Omega}{2\sqrt{3+2\omega}}  \frac{h}{m_p} 
    g^{\mu\nu} \partial_\mu  \xi \partial_\nu h    \left (1- \frac{\sqrt{2}}{ 2 \sqrt{3+2\omega}} \frac{\xi}{m_p}
      +  \frac{1}{4 (3+2\omega)} \frac{\xi^2}{m^2_p} \right )  \right .
    \\  &&  \nonumber \left .
     - 4   \lambda v^2  h^2  \Omega^{2} 
       \left (1- \frac{\sqrt{2}}{  \sqrt{3+2\omega}} \frac{\xi}{m_p}
      +  \frac{1}{(3+2\omega)} \frac{\xi^2}{m^2_p} \right )  \right .
     \\  &&  \nonumber 
     \left .
     - 4 \lambda v h^3
      \Omega
       \left (1- \frac{\sqrt{2}}{2 \sqrt{3+2\omega}} \frac{\xi}{m_p}
      +  \frac{1}{4(3+2\omega)} \frac{\xi^2}{m^2_p} \right )  
        -  \lambda    h^4  \right ) + {\cal O}\left(\frac{\xi}{m_p}\right)^3.
    \end{eqnarray}
Note that  fermion masses which are generated through  Yukawa couplings to the Higgs field are suppressed as well by the tiny exponential factor. The couplings of the Higgs field to the fermions are those of the standard model. We obtain new interactions between the fluctuations of the JBD-scalar field and the fermions of the standard model which are however Planck scale suppressed.

Furthermore, as in the Randall-Sundrum scenario we need to fine-tune the cosmological constant:
     \begin{eqnarray}
\Lambda_{eff}=   m_{\hat \phi}^2  m_p^2   \Omega^2+ 2 \lambda_{\hat \phi} m_p^4 +  \left(\Lambda - \frac{\lambda}{2} v^4\right) \Omega^4 
\end{eqnarray}
which implies a fine tuning of the dimensionless parameter $\lambda_{\hat \phi} $ which given our assumption on $m_{\hat \phi}$,  $m_p$ and $\Lambda$ needs to be chosen very small i.e. of the order of $-5 \times 10^{-52}$. Furthermore, as discussed previously, the mass of the JBD-field needs to be adjusted as well to reproduce the Planck scale. Note that these fine-tunings might be cured by quantum gravitational effects which are beyond the scope of our model.

If these two conditions are fulfilled the hierarchy problem of the standard model is solved and although there is a new scalar field in the low energy regime it is easy to see that its couplings to the fields of the standard model are suppressed by the Planck scale and it thus decouples from the standard model. Note that the weak scale is fixed by 
 \begin{eqnarray}
v_0=  {\left( -\frac{m_{\hat \phi}^2 m_p^2}{2 \Lambda - \lambda v} \right)} v \sim 10^{-17} v\sim 10^{-2}  \ \mbox{GeV}
\end{eqnarray}
  with $v\sim 10^{19}$ GeV. In other words we see that although the vacuum expectation value of the standard model Higgs field is big due to radiative corrections, its effective value is suppressed by the tiny exponential factor.   It is important to notice that in the Jordan frame the Planck scale does not exist and hence there is no hierarchy problem as the only scale of the model is the weak scale. This is not surprising since the notion of distance and hence energy scale is frame dependent since the invariant space-time interval $ds^2 =g_{\mu\nu}(x) dx^\mu dx^\nu$ is not invariant under the Weyl transformation we performed to go from the Jordan frame to the Einstein frame. 
   Furthermore the Einstein frame is more appropriate  for a discussion of the gauge hierarchy problem since in that frame we have a fundamental scale, i.e. the Planck mass, to which we can compare the weak scale.
In other words, when one looks at the theory in the Einstein frame, the Planck scale is reintroduced and the usual hierarchy problem reappears.  However,  we have shown that it is solved within our model by a rescaling of the Higgs field which redefines the weak scale and hence explains the splitting between the Planck scale and the weak scale.   The physical reason for the mismatch between the weak scale and the Planck scale is a violation of the Einstein's equivalence principle since in the Einstein frame the vacuum expectation value of the Higgs field is space-time dependent as can be seen from equation (\ref{fundaction}) i.e. $v(x)= \Omega(x) v$.

    Note that our model is closely related to an old idea by Fujii \cite{Fujii:1974bq}, Minkowski \cite{Minkowski:1977aj}  and   Zee \cite{Zee:1978wi} who proposed to generate the reduced Planck scale through a spontaneous symmetry breaking mechanism. Indeed we could think of generating the Planck scale using the JBD-field before doing the mapping to the Einstein frame, however our result would not be affected.   The new aspect of our work is to explain the gauge hierarchy of the standard model. A variation of the Zee model was proposed by van der Bij \cite{vanderBij:1994bv}, however in this model the Higgs boson couples only gravitationally to matter and hence decouples from the model. Let us also mention the work of Cognola et al. \cite{Cognola:2006eg} where the connection between a Gauss-Bonnet modification of gravity and the hierarchy problem has been considered. In our case the low energy theory is the usual standard model without new physics at the weak scale in the sense that the JBD-scalar field interacts only gravitationally with the standard model. It decouples completely from the standard model and we could have a naturally light Higgs mass stabilized by some Planck scale physics with a grand desert between the electroweak breaking scale and the Planck scale. The JBD-scalar field is very light and might be stable if its mass is smaller than that of the neutrinos. If this is the case it might be a viable dark matter candidate.  In any case, given its very weak couplings to the fields of the standard model it will be very difficult to discover this new particle. It is however interesting to see how the gauge hierarchy problem can be solved by new physics at the Planck scale.

 {\it Acknowledgments:} The author would like to thank Stephen Hsu and Archil Kobakhidze for helpful comments on this draft. He would also like to thank Jean-Marie Fr\`ere, Jean-Marc G\'erard and Thomas Hambye for helpful discussions.  This work was supported in part by the IISN and the  Belgian science policy office (IAP V/27).

\end{document}